\documentclass{emulateapj}

\usepackage{amsmath}

\begin{document}

\title{Constraining the Environment of CH$^+$ Formation with CH$_3^+$ Observations}

\author{Nick Indriolo\altaffilmark{1},
Takeshi Oka\altaffilmark{2},
T. R. Geballe\altaffilmark{3},
Benjamin J. McCall\altaffilmark{1,4}}

\altaffiltext{1}{Department of Astronomy, University of Illinois at
Urbana-Champaign, Urbana, IL 61801}
\altaffiltext{2}{Department of Astronomy and Astrophysics and Department of Chemistry, University of Chicago, Chicago, IL 60637}
\altaffiltext{3}{Gemini Observatory, 670 North A'ohoku Place, Hilo, HI 96720}
\altaffiltext{4}{Department of Chemistry, University of Illinois at
Urbana-Champaign, Urbana, IL 61801}

\begin{abstract}
The formation of CH$^+$ in the interstellar medium has long been an outstanding problem in chemical models.  In order to probe the physical conditions of the ISM in which CH$^+$ forms, we propose the use of CH$_3^+$ observations.  The pathway to forming CH$_3^+$ begins with CH$^+$, and a steady state analysis of CH$_3^+$ and the reaction intermediary CH$_2^+$ results in a relationship between the CH$^+$ and CH$_3^+$ abundances.  This relationship depends on the molecular hydrogen fraction, $f_{\rm H_2}$, and gas temperature, $T$, so observations of CH$^+$ and CH$_3^+$ can be used to infer the properties of the gas in which both species reside.  We present observations of both molecules along the diffuse cloud sight line toward Cyg OB2 No. 12.  Using our computed column densities and upper limits, we put constraints on the $f_{\rm H_2}$ vs. $T$ parameter space in which CH$^+$ and CH$_3^+$ form.  We find that average, static, diffuse molecular cloud conditions (i.e. $f_{\rm H_2}\gtrsim0.2$, $T\sim60$ K) are excluded by our analysis.  However, current theory suggests that non-equilibrium effects drive the reaction ${\rm C}^+ + {\rm H}_2 \rightarrow {\rm CH}^+ + {\rm H}$, endothermic by 4640 K.  If we consider a higher effective temperature due to collisions between neutrals and accelerated ions, the CH$_3^+$ partition function predicts that the overall population will be spread out into several excited rotational levels.  As a result, observations of more CH$_3^+$ transitions with higher signal-to-noise ratios are necessary to place any constraints on models where magnetic acceleration of ions drives the formation of CH$^+$.

\end{abstract}


\section{INTRODUCTION}

\subsection{Background}

Although CH$^+$ was first discovered in interstellar space nearly 70 years ago \citep{dunham37,douglas41}, the mechanism by which this simple molecule forms has remained elusive.  This is because many theoretical chemical models have been unable to reproduce the large observed abundance of CH$^+$ in diffuse cloud sight lines.  At present, it is thought that CH$^+$ is primarily formed by the reaction
\begin{equation}
{\rm C}^+ + {\rm H}_2 \rightarrow {\rm CH}^+ + {\rm H}.
\label{reach2pluscp}
\end{equation}
However, this reaction is highly endothermic \citep[$k_{\ref{reach2pluscp}}=1.0\times10^{-10}\exp{(-4640/T)}$~cm$^{3}$~s$^{-1}$;][]{federman96}, such that non-equilibrium chemistry must be invoked in order for it to proceed rapidly enough to produce the observed amounts of CH$^+$, without vastly overproducing observed abundances of OH.

Over the past several decades, various theories have been proposed to account for an increased rate of reaction (\ref{reach2pluscp}), including, but not limited to, neutral shocks \citep[e.g.][]{elitzur78,elitzur80}, magnetohydrodynamic (MHD) shocks \citep[e.g.][]{draine86}, Alfv\'{e}n waves \citep{federman96}, and turbulent dissipation \citep[e.g.][]{duley92,godard09,pan09}.  While both neutral and MHD shocks seem to have been ruled out by observations \citep{gredel93,crawford95}, turbulent dissipation and Alfv\'{e}n waves in diffuse clouds
remain viable mechanisms by which the rate of reaction (\ref{reach2pluscp}) may be increased, and there is no clear reason to favor one theory over the other at present.

\subsection{{\rm CH}$_3^+$ Chemistry}
To place constraints on the physical conditions of the interstellar medium (ISM) where CH$^+$ forms --- and potentially discriminate between the various proposed formation mechanisms --- we have investigated the reaction network linking CH$^+$ and CH$_3^+$, and made/obtained observations searching for both species.  In diffuse molecular clouds the carbon chemistry is either initiated by reaction (\ref{reach2pluscp}), or by the radiative association reaction
\begin{equation}
{\rm C^+} + {\rm H_2} \rightarrow {\rm CH_2^+} + h\nu,
\label{reach2cphv}
\end{equation}
which ``bypasses'' CH$^+$ production.  The branching fraction between these reactions is temperature dependent, and reaction (\ref{reach2pluscp}) will dominate when collision temperatures are greater than about 400~K.  Assuming CH$^+$ {\it is} formed, it will react with molecular hydrogen,
\begin{equation}
{\rm CH^+} + {\rm H_2} \rightarrow {\rm CH_2^+} + {\rm H},
\label{reacchplush2}
\end{equation}
and eventually form CH$_3^+$ via
\begin{equation}
{\rm CH_2^+} + {\rm H_2} \rightarrow {\rm CH_3^+} + {\rm H}.
\label{reacch2plush2}
\end{equation}
In addition to reaction (\ref{reacch2plush2}) though, the CH$_2^+$ intermediary is also destroyed by dissociative recombination with electrons
\begin{equation}
{\rm CH_2^+} + e^- \rightarrow {\rm products},
\label{reacch2pluseall}
\end{equation}
thus decreasing the CH$_3^+$ production rate.  Dissociative recombination with electrons also happens to be the primary process by which CH$_3^+$ is destroyed:
\begin{equation}
{\rm CH_3^+} + e^- \rightarrow {\rm products}
\label{reacch3pluseall}
\end{equation}
(there are multiple product channels for reactions (\ref{reacch2pluseall}) and (\ref{reacch3pluseall}), but our analysis only depends on the overall dissociative recombination rates).  While CH$_3^+$ will be formed starting from both reactions (\ref{reach2cphv}) and (\ref{reacchplush2}), the latter process will dominate in clouds containing CH$^+$ ($k_{\ref{reacchplush2}}x({\rm CH}^+)/k_{\ref{reach2cphv}}x({\rm C}^+)\gtrsim100$, where $k_{\ref{reacchplush2}}=1.2\times10^{-9}$~cm$^3$~s$^{-1}$ \citep{mcewan99}, $k_{\ref{reach2cphv}}=4\times10^{-16}(T/300)^{-0.2}$~cm$^3$~s$^{-1}$ \citep{herbst82,herbst85}, $x({\rm CH}^+)\sim10^{-8}$ \citep{sheffer08}, and $x({\rm C}^+)\sim 1.4\times10^{-4}$ \citep{cardelli96}).  Because we are studying cloud components with CH$^+$, we omit reaction (\ref{reach2cphv}) from our analysis.  Note, though, that if CH$_3^+$ were observed in a cloud component lacking CH$^+$, the following analysis would not be applicable.

Assuming steady state for both CH$_2^+$ and CH$_3^+$, we can derive a relation between the concentrations of CH$_3^+$ and CH$^+$, given by
\begin{equation}
\frac{n({\rm CH_3^+})}{n({\rm CH^+})}=\frac{f_{\rm H_2}^2k_{\ref{reacchplush2}}}{2k_{\ref{reacch3pluseall}}x_e(f_{\rm H_2}+2x_ek_{\ref{reacch2pluseall}}/k_{\ref{reacch2plush2}})}.
\label{eqch3chratio}
\end{equation}
Here, the $k_i$'s are the rate coefficients for reaction $i$,
$f_{\rm H_2}\equiv 2n({\rm H_2})/n_{\rm H}$ is the molecular hydrogen fraction (where $n_{\rm H}\equiv n({\rm H})+2n({\rm H_2})$), and $x_e$ is the electron fraction ($x_e\equiv n_e/n_{\rm H}$).  The rate coefficients used in this study are 
$k_{\ref{reacch2plush2}}=1.6\times10^{-9}$~cm$^3$~s$^{-1}$ \citep{smith77}, $k_{\ref{reacch2pluseall}}=6.4\times10^{-7}(T/300)^{-0.6}$~cm$^3$~s$^{-1}$ \citep{larson98}, and $k_{\ref{reacch3pluseall}}=3.5\times10^{-7}(T/300)^{-0.5}$~cm$^3$~s$^{-1}$ \citep{mitchell90}.  However, \citet{sheehan04} note that $k_{\ref{reacch3pluseall}}$ was determined using vibrationally excited CH$_3^+$.  The rate coefficient for dissociative recombination from the ground vibrational state --- the only state likely to be populated in diffuse cloud conditions --- may differ from the above experimental value by about a factor of 3.  Because of this possibility, during our analysis we also consider using $3k_{\ref{reacch3pluseall}}$ and $k_{\ref{reacch3pluseall}}/3$ in equation (\ref{eqch3chratio}).

Aside from the electron fraction --- estimated to be $x_e\sim 1.4\times10^{-4}$ in diffuse clouds via C$^+$ observations \citep{cardelli96} --- equation (\ref{eqch3chratio}) is dependent only on the molecular hydrogen fraction and the temperature (through the rate coefficients $k_{\ref{reacch2pluseall}}$ and $k_{\ref{reacch3pluseall}}$).  As a result, the ratio between the abundances of CH$_3^+$ and CH$^+$ is governed by the interstellar physical parameters $f_{\rm H_2}$ and $T$.  Observations of CH$^+$ and CH$_3^+$ can thus be used to infer the physical conditions in the ISM where these molecules reside.

\section{OBSERVATIONS \& DATA REDUCTION}

\subsection{The Cyg OB2 No. 12 Sight Line}

We selected the Cyg OB2 No. 12 sight line for this study because it combines a large extinction \citep[$A_V\sim10$;][]{schulte58} and thus gas column, with a background star that is bright enough at both optical and infrared wavelengths such that the observations described below are possible.  However, there is some contention as to whether the sight line mainly probes diffuse or dense material, environments which have different electron fractions.  Rotational excitation analyses have been performed using C$_2$ observations, and the resulting inferred hydrogen number densities are between $200$~cm$^{-3}$ and $400$~cm$^{-3}$ \citep{gredel01,sonnentrucker07}, typical of diffuse clouds.  Also, the non-detection of H$_2$O and CO$_2$ ices by \citet{whittet97} led to the conclusion that there was a lack of dense molecular gas along the sight line.  On the other hand, radio observations of HCO$^+$ \citep{scappini00} and $^{13}$CO \citep{casu05} suggest that portions of the sight line may pass through denser clumps as proposed by \citet{cecchipestellini00}.  Additionally, \citet{cecchipestellini02} show that dense clumps can still be present even with the aforementioned C$_2$ analyses.  Because we are examining the chemistry related to CH$^+$, a species thought to reside mainly in diffuse gas with $10~{\rm cm}^{-3}<n_{\rm H}<300~{\rm cm}^{-3}$ \citep{pan05}, we adopt the diffuse molecular cloud model for our analysis.  However, in Section 4 we also consider a reduced electron fraction in order to investigate the effects of varying this parameter.

\subsection{${\rm CH}_3^+$}

Observations toward Cyg OB2 No. 12 and the atmospheric standard $\alpha$~Cyg were made on 1999 Nov 19 using the CGS4 spectrometer \citep{mountain90} at UKIRT.  The spectrometer was used with its echelle grating, 0.6'' wide slit, and long camera to yield a resolving power of about 37,000, and a circular variable filter (CVF) was employed to select the correct order.  Spectra were centered to cover the $^rR(1,0)$, $^rR(1,1)$, and $^rR(2,2)$ ro-vibrational transitions of CH$_3^+$ at 3.18523~$\mu$m, 3.19578~$\mu$m, and 3.18777~$\mu$m (vacuum wavelengths), respectively \citep{crofton85}.  Stars were nodded along the slit in an ABBA pattern with total integration times of 864~s for Cyg OB2 No. 12 and 672~s for $\alpha$~Cyg.

\begin{figure}
\epsscale{1.25}
\plotone{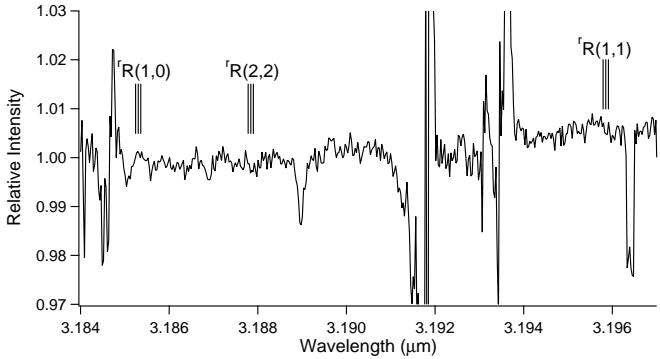}
\caption{This spectrum of Cyg OB2 No. 12 covers the $^rR(1,0)$, $^rR(1,1)$, and $^rR(2,2)$ transitions of CH$_3^+$.  The three vertical lines underneath each transition label show the expected positions for absorption due to various cloud components along this line of sight.  The $\gtrsim2\%$ fluctuations near 3.185~$\mu$m, 3.1965~$\mu$m, and between 3.191~$\mu$m and 3.194~$\mu$m are due to strong telluric absorption features which could not be completely removed through standard star division.  The feature near 3.189~$\mu$m is an instrumental artifact.}
\label{figch3plus}
\end{figure}

Individual frames were processed using Starlink's ORAC-DR pipeline\footnote{See web site at http://www.oracdr.org/.}, specifically designed to handle UKIRT data. Spectra were then extracted from the resultant group frame (the combination of all integrations for a given target) using NOAO's IRAF package\footnote{See web site at http://iraf.noao.edu/.}, and imported to IGOR Pro\footnote{See web site at http://www.wavemetrics.com/.} where we have macros set up to complete the reduction \citep{mccallthesis}.  Here, artifacts from the data acquisition methods were removed, spectra were wavelength calibrated using telluric lines, and the science target spectra were divided by the telluric standard spectra to remove atmospheric absorption features.  The resulting normalized spectrum of Cyg OB2 No. 12 is shown in Figure \ref{figch3plus}.

\subsection{${\rm CH}^+$ and {\rm CH}}

The $A$--$X$(0-0) and $A$--$X$(1-0) transitions of CH$^+$ \citep[4232.548~\AA\ and 3957.692~\AA, respectively;][]{carrington82} and the $A$--$X$(0-0) transition of CH \citep[4300.308~\AA;][]{zachwieja95}, were observed simultaneously using the HIRES instrument at Keck \citep{vogt94} by G. Blake and collaborators.  Observations toward Cyg OB2 No. 12 were performed on 1999 Jun 17 and 18 with a slit width of 0.57'', producing a resolving power of about 70,000.  The total integration time on source was 8400~s.

\begin{figure}
\epsscale{1.25}
\plotone{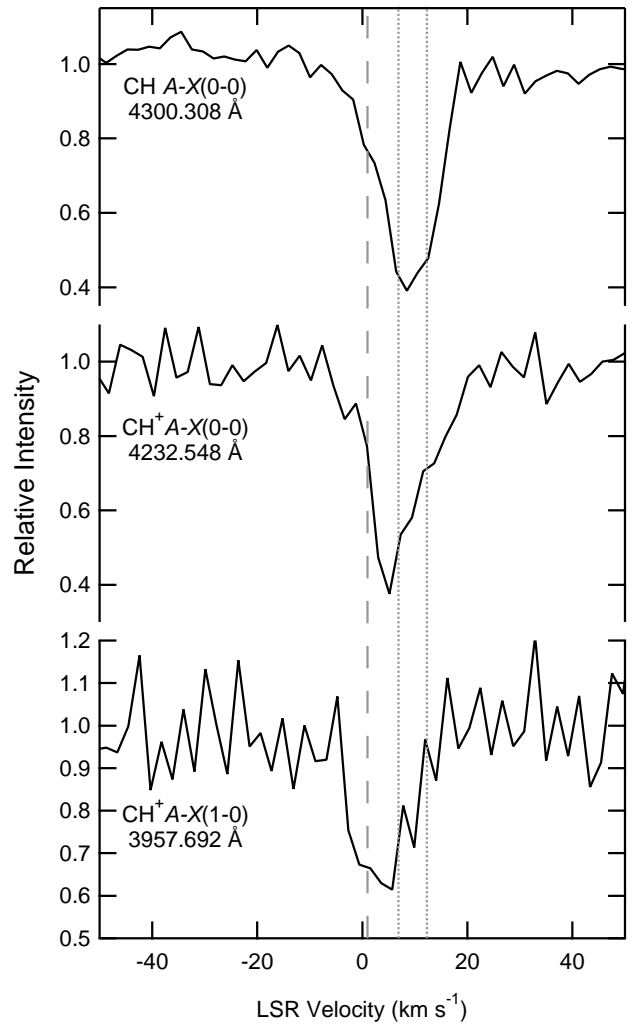}
\caption{Spectra of Cyg OB2 No. 12 showing the $A$--$X$(0-0) transition of CH (top spectrum) and the $A$--$X$(0-0) and $A$--$X$(1-0) transitions of CH$^+$ (middle and bottom spectra, respectively). The vertical dotted lines mark cloud components at 6.9~km~s$^{-1}$ and 12.3~km~s$^{-1}$ seen by \citet{mccall02} in CN, C$_2$, $^{12}$CO, H$_3^+$, and K \textsc{i}.  The dashed line marks a 1~km~s$^{-1}$ component only seen in H$_3^+$ and K \textsc{i}.}
\label{figchplus}
\end{figure}

Using IRAF, the profile of the overscan region was subtracted from each image, and the frames were averaged together with the {\it cosmic ray reject} option enabled.
The relevant orders (10, 16, \& 17) were extracted using {\it apall} with background subtraction enabled\footnote{The initial background subtraction for order 10, which contains the 3957.692~\AA\ line of CH$^+$, resulted in intensities below zero at the centers of Ca~\textsc{ii} lines near 3934 and 3968~\AA.  This is most likely due to the low flux and low signal-to-noise ratio of this order.  A curve of growth analysis (see Section 3.2) using the equivalent width of the 3957.692~\AA\ line from this uncorrected spectrum resulted in no value of the Doppler parameter for which the column densities determined from the $A$--$X$(0-0) and $A$--$X$(1-0) absorption lines agreed.  To remedy these unphysical results, we shifted the spectrum of order 10 so that the deeper Ca~\textsc{ii} line had zero intensity.}, and normalized using the {\it continuum} routine.  These astronomical spectra were then imported to IGOR PRO, along with Thorium-Argon arc lamp spectra used for wavelength calibration.  After calibration, observed wavelengths were converted to LSR velocities, and the resulting spectra are shown in Figure \ref{figchplus}.

\section{RESULTS}

\subsection{${\rm CH}_3^+$}

The spectrum in Figure \ref{figch3plus} shows no indication of absorption from any of the CH$_3^+$ transitions considered.  Upper limits on the equivalent widths were computed using $W_{\lambda}<\sigma\lambda_{\rm pix}\sqrt{{\cal N}_{\rm pix}}$, where $\sigma$ is the standard deviation on the continuum of the spectrum, $\lambda_{\rm pix}$ is the wavelength per pixel, and ${\cal N}_{\rm pix}$ is the number of pixels expected in an absorption feature given a full width at half-maximum (FWHM) of 12.4~km~s$^{-1}$ (average FWHM measured for the CH$^+$ and CH lines).  The upper limits on column densities were then calculated using the standard relation between equivalent width and column.  The results of this analysis are shown in Table \ref{tblch3plus}.

\subsection{${\rm CH}^+$ and {\rm CH}}

The CH$^+$ and CH spectra in Figure \ref{figchplus} show broad and deep absorption profiles.  While it is known that there are multiple velocity components toward Cyg OB2 No. 12, the low signal-to-noise ratios of the CH and CH$^+$ spectra make fitting multiple components to the observed absorption features highly uncertain.  Because there is no unique way to fit any of the absorption features with 3 separate components, we simply found the equivalent widths of each feature as a whole.  Using these equivalent widths, we first computed lower limits to the column densities of both species by assuming that the lines are optically thin.  As this assumption is most likely incorrect, we also computed column densities using a curve of growth analysis.  Both CH$^+$ lines arise from the same lower energy state, and so should result in the same column density.  Plotting the column density vs. the Doppler parameter, $b$, for both lines, we found that the curves intersect at $b=3.5_{+4}^{-1}$~km~s$^{-1}$ and $N({\rm CH^+})=2.2^{+1.5}_{-0.6}\times10^{14}$~cm$^{-2}$ within the uncertainties of the equivalent widths (the positive uncertainty associated with $b$ was computed from the average FWHM of the CH$^+$ lines and the relation $b=\Delta v_{\rm FWHM}/(2\sqrt{\ln(2)}$)). Assuming that $b=3.5_{+4}^{-1}$~km~s$^{-1}$ is also applicable to CH, we computed $N({\rm CH})=2.8_{-1}^{+4}\times10^{14}$~cm$^{-2}$.  All of these results, as well as the various line parameters, are shown in Table \ref{tblchplus}.

\subsection{Interstellar Conditions}

\begin{figure}
\epsscale{1.25}
\plotone{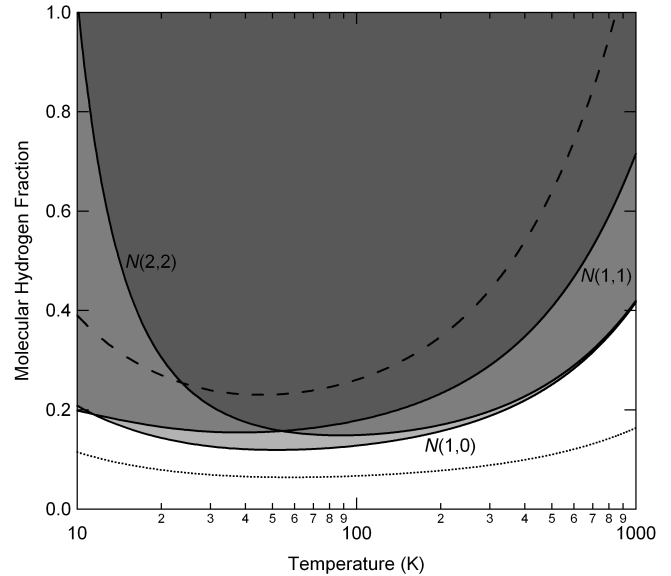}
\caption{Plot of the ($f_{\rm H_2},T$) parameter space.  The 3 solid curves represent contours corresponding to the values of $f_{\rm H_2}$ and $T$ which reproduce the ratios $N(1,0)/N({\rm CH^+})$, $N(1,1)/N({\rm CH^+})$, and $N(2,2)/N({\rm CH^+})$ (where the $N(J,K)$ values are $3\sigma$ upper limits) using equation (\ref{eqch3Jkchratio}), and are labeled accordingly.  Shaded regions are excluded by our analysis, and progressively darker shading indicates regions excluded by more than one observation.  The dashed and dotted curves show the parameter space excluded by the $N(1,0)$ upper limit using $3k_{\ref{reacch3pluseall}}$ and $k_{\ref{reacch3pluseall}}/3$, respectively.}
\label{figfTexclude}
\end{figure}

Using both CH$^+$ and CH$_3^+$ observations, and assuming that the two species have the same distribution along the line of sight (i.e. $N({\rm CH}_3^+)/N({\rm CH}^+)=n({\rm CH}_3^+)/n({\rm CH}^+)$), equation (\ref{eqch3chratio}) can be used to constrain the temperature and molecular fraction of the gas where these species reside.  However, equation (\ref{eqch3chratio}) requires the {\it total} abundance of CH$_3^+$, whereas we only have upper limits for the column densities of three particular states.  As a result, the equation must be recast such that the left hand side is the ratio between the column density of an individual state of CH$_3^+$ with the total column of CH$^+$.  This is achieved by using the partition function for CH$_3^+$ (computed using molecular constants from \citealt{crofton88} and \citealt{jagod94}) to calculate the fractional population, $P_{J,K}(T)$, in each individual state as a function of temperature.  The definition $N({\rm CH}_3^+)P_{J,K}(T)=N(J,K)$ is then substituted into equation (\ref{eqch3chratio}), giving
\begin{equation}
\frac{N(J,K)}{N({\rm CH^+})}=P_{J,K}(T)\frac{f_{\rm H_2}^2k_{\ref{reacchplush2}}}{2k_{\ref{reacch3pluseall}}x_e(f_{\rm H_2}+2x_ek_{\ref{reacch2pluseall}}/k_{\ref{reacch2plush2}})}.
\label{eqch3Jkchratio}
\end{equation}
Using the observed CH$^+$ column and the individual $3\sigma$ upper limits for CH$_3^+$, we can put three separate upper limits on the left hand side of the equation.  The ($f_{\rm H_2},T$) parameter space can then be explored to determine which molecular fraction--temperature combinations are excluded by our observations.  This analysis is shown in Figure \ref{figfTexclude}.

\section{DISCUSSION}

It is quite clear from Figure \ref{figfTexclude} that our analysis using CH$^+$ and CH$_3^+$ observations excludes a large portion of the ($f_{\rm H_2},T$) parameter space.  Indeed, typical diffuse molecular cloud conditions ($f_{\rm H_2}\gtrsim0.2$, $T\sim60$~K) are excluded by our upper limits to the column densities of all three observed CH$_3^+$ transitions.  While this may not be especially surprising --- equilibrium chemistry in diffuse clouds is unable to account for the large abundances of CH$^+$ --- our observations add a new, independent method for determining that CH$^+$ cannot form under such conditions.

However, as mentioned in Section 1.2, the dissociative recombination rate coefficient for CH$_3^+$ with electrons, $k_{\ref{reacch3pluseall}}$, is most likely uncertain by a factor of about 3.  If we allow variations between $3k_{\ref{reacch3pluseall}}$ and $k_{\ref{reacch3pluseall}}/3$ during our analysis using $N(1,0)$ with equation (\ref{eqch3Jkchratio}), the dashed and dotted contours in Figure \ref{figfTexclude} result, respectively.  With $k_{\ref{reacch3pluseall}}/3$ essentially every temperature is excluded when the molecular fraction is greater than 0.1.  For $3k_{\ref{reacch3pluseall}}$, high molecular fractions are allowed as the temperature increases beyond a few hundred Kelvin.  Even with this higher rate coefficient though, most typical diffuse molecular clouds conditions are still excluded by this analysis.

While we have assumed diffuse cloud conditions throughout our analysis and given reasons for doing so, we feel it is prudent to revisit the dense clump scenario described in Section 2.1.  For the sake of completeness, we examine the effects that a lower electron fraction (due to denser material) would have on our analysis.  If we decrease the electron fraction by a factor of 10 to $x_e=1.4\times10^{-5}$ (still much higher than $x_e\sim4\times10^{-8}$ assumed in dense clouds \citep{woodall07}), then interstellar clouds with $f_{\rm H_2}>0.04$ and $10~{\rm K} < T < 1000~{\rm K}$ are excluded.  This demonstrates that as $x_e$ decreases, regions with even smaller molecular hydrogen fractions are ruled out by our analysis.

A potentially larger uncertainty in our analysis is that we only consider a single kinetic temperature of the gas, and do not account for any non-equilibrium effects.  If ions are magnetically accelerated relative to neutrals, our analysis changes drastically.  Assuming an effective temperature for collisions can be described as \citep{flower85,federman96}
\begin{equation}
T_{\rm eff}=T_{\rm kin}+\frac{\mu}{3k_B}(\Delta v)^2,
\label{eqTeff}
\end{equation}
where $\mu$ is the reduced mass of collision partners, $k_B$ is Boltzmann's constant, and $\Delta v$ is the turbulent velocity \citep[about 3.3 km s$^{-1}$ accoring to][]{sheffer08}, the effective temperatures for collisions of CH$_3^+$ with H$_2$ and H are about 800~K and 400~K higher, respectively, than the kinetic temperature of the gas.
These higher temperatures are important because it is generally assumed that the fractional populations of CH$_3^+$ states are ``thermalized'' by collisions with H and H$_2$.  If collision temperatures are hundreds of Kelvin instead of $\sim$60~K, then the expected populations of the (1,0), (1,1), and (2,2) states we observed are significantly decreased.  In this case, our analysis is unable to exclude any significant portion of the ($f_{\rm H_2},T$) parameter space.

To use our method of analysis assuming high effective temperatures due to accelerated ions, new observations must be made.  Such observations should cover the 3 transitions we have already examined, search for absorption due to higher energy states of CH$_3^+$, and obtain a higher signal-to-noise ratio in all cases. If absorption from CH$_3^+$ can be detected, instead of only excluding portions of ($f_{\rm H_2},T$) space, we will be able to constrain the conditions where CH$^+$ {\it does} form.

\section{CONCLUSIONS}

We have used observations of CH$_3^+$ and CH$^+$ toward the diffuse molecular cloud sight line Cyg OB2 No. 12 in an attempt to constrain the physical conditions of the ISM where these species reside.  If we assume equilibrium chemistry, our analysis excludes the portion of ($f_{\rm H_2},T$) parameter space corresponding to average diffuse molecular clouds.  This acts as an important independent check on previous studies which also concluded that the slow rate of CH$^+$ formation ruled out equilibrium chemistry in such environments.  However, our analysis cannot exclude non-equilibrium effects in diffuse clouds, and neither supports nor refutes any of the current proposed theories for CH$^+$ formation.  To do so, new observations of CH$_3^+$ with signal-to-noise ratios much higher than those obtained in this study ($\sim600$) would be required.  Also, it would be highly advantageous if the dissociative recombination rate coefficient for CH$_3^+$ were measured at low vibrational temperatures, i.e. for the ground vibrational state.

\mbox{}

The authors would like to thank Geoff Blake, Ted Snow, Ralph Shuping, and Mike Brown for providing us with their HIRES observations of Cyg OB2 No. 12 and calibration frames, Dan Welty for assisting in the analysis of the CH and CH$^+$ data, and Steve Federman and the anonymous referee for helpful feedback.  N.I. and B.J.M. are supported by NSF grant PHY 08-55633.  T.O. is supported by NSF grant AST-0849577. T.R.G.'s research is supported by the Gemini Observatory, which is operated by the Association of Universities for Research in Astronomy, Inc., on behalf of the international Gemini partnership of Argentina, Australia, Brazil, Canada, Chile, the United Kingdom, and the United States of America.  The United Kingdom Infrared Telescope is operated by the Joint Astronomy Centre on behalf of the Science and Technology Facilities Council of the U.K.  Some of the data presented herein were obtained at the W.M. Keck Observatory, which is operated as a scientific partnership among the California Institute of Technology, the University of California and the National Aeronautics and Space Administration. The Observatory was made possible by the generous financial support of the W. M. Keck Foundation.

\clearpage

\begin{deluxetable}{cccc}

\tablecaption{CH$_3^+$ Absorption Line Parameters  \label{tblch3plus}}
\tablehead{ & \colhead{$|\mu|^2$} & \colhead{$W_{\lambda}$} & \colhead{$N_{\rm level}$} \\
\colhead{Transition} & \colhead{(D$^2$)} & \colhead{(m\AA)} & \colhead{($10^{12}$~cm$^{-2}$)}
}

\startdata
$^rR(1,0)$ & .0313 & $<1.34$ & $<3.23$  \\
$^rR(1,1)$ & .0313 & $<1.21$ & $<2.91$  \\
$^rR(2,2)$ & .0313 & $<1.10$ & $<2.65$
\enddata
\tablecomments{Values for $W_{\lambda}$ and $N_{\rm level}$ are $1\sigma$ upper limits.  Transition dipole moments were calculated using H\"{o}nl-London factors and the transition moment given by \citet{pracna93}}

\end{deluxetable}


\clearpage

\begin{deluxetable}{cccccccc}

\tablecaption{CH and CH$^+$ Absorption Line Parameters \label{tblchplus}}
\tablehead{ & & & \colhead{$W_{\lambda}$} & \colhead{FWHM} & \colhead{$N_{\rm thin}$} & \colhead{$N$} & \colhead{$b$} \\
\colhead{Species} & \colhead{Transition} & \colhead{$f$} & \colhead{(m\AA)} & \colhead{(km~s$^{-1}$)} & \colhead{($10^{14}$~cm$^{-2}$)} & \colhead{($10^{14}$~cm$^{-2}$)} & \colhead{(km~s$^{-1}$)}
}
\startdata
CH$^+$ & $A$--$X$(0-0) & 0.00545\tablenotemark{a} & $103\pm6$ & 13.7 & $>1.2$ & $2.2^{+1.5}_{-0.6}$ & 3.5$_{+4}^{-1}$ \\
CH$^+$ & $A$--$X$(1-0) & 0.00331\tablenotemark{a} & $~70\pm8$ & 11.3 & $>1.5$ & $2.2^{+1.5}_{-0.6}$ & 3.5$_{+4}^{-1}$ \\
CH     & $A$--$X$(0-0) & 0.00506\tablenotemark{b} & $113\pm4$ & 12.2 & $>1.4$ & $2.8^{+4}_{-1}$ & 3.5$_{+4}^{-1}$
\enddata

\tablecomments{~Lower limits ($N_{\rm thin}$) were computed under the assumption that the gas is optically thin.  As this is most likely not the case, we performed a curve of growth analysis for both CH$^+$ lines (which arise from the same lower state) in order to find the $b$-value where both lines predict the same CH$^+$ column density.  This occurs at $b=3.5_{+4}^{-1}$~km~s$^{-1}$, where $N({\rm CH^+})=2.2^{+1.5}_{-0.6}\times10^{14}$~cm$^{-2}$.}
\tablenotetext{a}{Oscillator strength from \citet{larsson83a}.}
\tablenotetext{b}{Oscillator strength from \citet{larsson83b}.}

\end{deluxetable}



\begin{thebibliography}{}
\bibitem[Cardelli et al.(1996)]{cardelli96} Cardelli, J. A., Meyer, D. M., Jura, M., \& Savage, B. D. 1996, \apj, 467, 334
\bibitem[Carrington \& Ramsay(1982)]{carrington82} Carrington, A., \& Ramsay, D. A. 1982, Physica Scripta, 25, 272
\bibitem[Casu et al.(2005)]{casu05} Casu, S., Scappini, F., Cecchi-Pestellini, C., \& Olberg, M. 2005, \mnras, 359, 73
\bibitem[Cecchi-Pestellini \& Dalgarno(2000)]{cecchipestellini00} Cecchi-Pestellini, C., \& Dalgarno, A. 2000, \mnras, 313, L6
\bibitem[Cecchi-Pestellini \& Dalgarno(2002)]{cecchipestellini02} Cecchi-Pestellini, C., \& Dalgarno, A. 2002, \mnras, 331, L31
\bibitem[Crawford(1995)]{crawford95} Crawford, I. A. 1995, \mnras, 277, 458
\bibitem[Crofton et al.(1988)]{crofton88} Crofton, M. W., Jagod, M.-F., Rehfuss, B. D., Kreiner, W. A., \& Oka, T. 1988, J. Chem. Phys., 88, 666
\bibitem[Crofton et al.(1985)]{crofton85} Crofton, M. W., Kreiner, W. A., Jagod, M.-F., Rehfuss, B. D., \& Oka, T. 1985, J. Chem. Phys., 83, 3702
\bibitem[Douglas \& Herzberg(1941)]{douglas41} Douglas, A. E., \& Herzberg, G. 1941, \apj, 94, 381
\bibitem[Draine \& Katz(1986)]{draine86} Draine, B. T., \& Katz, N. 1986, \apj, 310, 392
\bibitem[Duley et al.(1992)]{duley92} Duley, W. W., Hartquist, T. W., Sternberg, A., Wagenblast, R., \& Williams, D. A. 1992, \mnras, 255, 463
\bibitem[Dunham(1937)]{dunham37} Dunham, T., Jr. 1937, PASP, 49, 26
\bibitem[Elitzur \& Watson(1978)]{elitzur78} Elitzur, M., \& Watson, W. D. 1978, \apj, 222, L141
\bibitem[Elitzur \& Watson(1980)]{elitzur80} Elitzur, M., \& Watson, W. D. 1980, \apj, 236, 172
\bibitem[Federman et al.(1996)]{federman96} Federman, S. R., Rawlings, J. M. C., Taylor, S. D., \& Williams, D. A. 1996, \mnras, 279, L41
\bibitem[Flower et al.(1985)]{flower85} Flower, D. R., Pineau des Forets, G., \& Hartquist, T. W. 1985, \mnras, 216, 775
\bibitem[Godard et al.(2009)]{godard09} Godard, B., Falgarone, E., \& Pineau des For\^{e}ts, G. 2009, \aap, 495, 847
\bibitem[Gredel et al.(2001)]{gredel01} Gredel, R., Black, J. H., \& Yan, M. 2001, \aap, 375, 553
\bibitem[Gredel et al.(1993)]{gredel93} Gredel, R., van Dishoeck, E. F., \& Black, J. H. 1993, \aap, 269, 477
\bibitem[Herbst(1982)]{herbst82} Herbst, E. 1982, \apj, 252, 810
\bibitem[Herbst(1985)]{herbst85} Herbst, E. 1985, \apj, 291, 226
\bibitem[Jagod et al.(1994)]{jagod94} Jagod, M.-F., Gabrys, C. M., R\"{o}sslein, M., Uy, D., \& Oka, T. 1994, Can. J. Phys., 72, 1192
\bibitem[Larson et al.(1998)]{larson98} Larson, \AA., et al. 1998, \apj, 505, 459
\bibitem[Larsson \& Siegbahn(1983a)]{larsson83a} Larsson, M., \& Siegbahn, P. E. M. 1983a, Chem. Phys., 76, 175
\bibitem[Larsson \& Siegbahn(1983b)]{larsson83b} Larsson, M., \& Siegbahn, P. E. M. 1983b, J. Chem. Phys. 79, 2270; 85, 4208
\bibitem[McCall(2001)]{mccallthesis} McCall, B. J. 2001, Ph.D. thesis, Univ. Chicago
\bibitem[McCall et al.(2002)]{mccall02} McCall, B. J., et al. 2002, \apj, 567, 391
\bibitem[McEwan et al.(1999)]{mcewan99} McEwan, M. J., Scott, G. B. I., Adams, N. G., Babcock, L. M., Terzieva, R., \& Herbst, E. \apj, 513, 287
\bibitem[Mitchell(1990)]{mitchell90} Mitchell, J. B. A. 1990, Phys. Rep., 186, 215
\bibitem[Mountain et al.(1990)]{mountain90} Mountain, C. M., Robertson, D. J., Lee, T. J., \& Wade, R. 1990, Proc. SPIE, 1235, 25
\bibitem[Pan et al.(2005)]{pan05} Pan, K., Federman, S. R., Sheffer, Y., \& Andersson, B.-G. 2005, \apj, 633, 986
\bibitem[Pan \& Padoan(2009)]{pan09} Pan, L., \& Padoan, P. 2009, \apj, 692, 594
\bibitem[Pracna et al.(1993)]{pracna93} Pracna, P., \u{S}pirko, V., \& Kraemer, W. P. 1993, J. Mol. Spectrosc., 158, 433
\bibitem[Scappini et al.(2000)]{scappini00} Scappini, F., Cecchi-Pestellini, C., Codella, C., \& Dalgarno, A. 2000, \mnras, 317, L6
\bibitem[Schulte(1958)]{schulte58} Schulte, D. H. 1958, \apj, 128, 41
\bibitem[Sheehan \& St.-Maurice(2004)]{sheehan04} Sheehan, C. H., \& St.-Maurice, J.-P. 2004, Adv. Space Res., 33, 216
\bibitem[Sheffer et al.(2008)]{sheffer08} Sheffer, Y., Rogers, M., Federman, S. R., Abel, N. P., Gredel, R., Lambert, D. L., \& Shaw, G. 2008, \apj, 687, 1075
\bibitem[Smith \& Adams(1977)]{smith77} Smith, D., \& Adams, N. G. 1977, Chem. Phys. Let., 47, 383
\bibitem[Sonnentrucker et al.(2007)]{sonnentrucker07} Sonnentrucker, P., Welty, D. E., Thorburn, J. A., \& York, D. G. 2007, \apjs, 168, 58
\bibitem[Vogt et al.(1994)]{vogt94} Vogt, S. S., et al. 1994, Proc. SPIE, 2198, 362
\bibitem[Whittet et al.(1997)]{whittet97} Whittet, D. C. B., et al. 1997, \apj, 490, 729
\bibitem[Woodall et al.(2007)]{woodall07} Woodall, J., Ag\'{u}ndez, M., Markwick-Kemper, A. J., \& Millar,~T.~J. 2007, \aap, 466, 1197
\bibitem[Zachwieja(1995)]{zachwieja95} Zachwieja, M. 1995, J. Mol. Spectrosc., 170, 285
\end{thebibliography}
\end{document}